\documentclass{article}

\usepackage{arxiv}

\usepackage[utf8]{inputenc} 
\usepackage[T1]{fontenc}    
\usepackage{amsmath}
\usepackage{amssymb}
\usepackage{amsfonts}       
\usepackage{amsthm}
\usepackage{url}            
\usepackage{booktabs}       
\usepackage{nicefrac}       
\usepackage{microtype}      
\usepackage{graphicx}
\usepackage{doi}
\usepackage{algorithm}
\usepackage{algpseudocode}
\usepackage{tabularx}
\usepackage{array}
\usepackage{multirow}
\usepackage{enumitem}
\usepackage{cleveref}       
\usepackage{xcolor}
\usepackage{hyperref}   
\hypersetup{colorlinks=true, citecolor=blue}

\theoremstyle{plain}
\newtheorem{proposition}{Proposition}[section]
\newtheorem{corollary}{Corollary}[section]
\newtheorem{assumption}{Assumption}[section]
\theoremstyle{definition}
\newtheorem{definition}{Definition}[section]

\numberwithin{equation}{section}

\title{\textnormal{StateGate: A Behavioral-Fidelity Prompt Optimization Method for Long-Chain LLM Agents}}

\newif\ifuniqueAffiliation
\uniqueAffiliationtrue

\ifuniqueAffiliation
\author{
	Shicheng Huang\textsuperscript{1} \and
	Cai Fu\textsuperscript{1, *} \and
	Jianqiang Lyu\textsuperscript{2} \and
	Yifan Zhang\textsuperscript{1} \and
	Xianmao Ji\textsuperscript{1}
}
\else
\fi

\renewcommand{\shorttitle}{\textit{StateGate}: Behavioral-Fidelity Prompt Optimization}

\hypersetup{
hidelinks,
hypertexnames=false,
pdftitle={StateGate: A Behavioral-Fidelity Prompt Optimization Method for Long-Chain LLM Agents},
pdfsubject={cs.CR, cs.AI},
pdfauthor={Anonymous Author(s)},
pdfkeywords={LLM agents, semantic constraints, symbolic constraints, behavioral fidelity, vulnerability discovery},
}

\begin{document}
\maketitle

\vspace{-1em}
\noindent\footnotesize
\textsuperscript{1} School of Cyber Science and Engineering, Huazhong University of Science and Technology, Wuhan 430074, China \\
\textsuperscript{2} School of Cyberspace Security, Cyberspace Security University of China \\
\textsuperscript{*} Corresponding author: Cai Fu (e-mail: fucai@hust.edu.cn).
\normalsize
\vspace{1em}

\begin{abstract}
	Currently, long-chain LLM agents commonly suffer from behavioral degradation in complex tasks, where prompts are written in full but steps are not executed in full. To address this issue, existing approaches face three major challenges. First, constraints are strongly bound to their textual carriers; once natural-language constraint sentences are removed, their binding force vanishes. Second, the execution process lacks independent verification, as the evidence for judgment is often supplied by the executing party itself, leading to the failure of self-referential validation. Third, existing remediation schemes are difficult to generalize, and hard-coded mechanisms cannot be transferred across tasks. In this paper, we propose StateGate, a novel prompt-optimization scheme that transforms semantic constraints into symbolic gates enforced by an external deterministic runtime. By decoupling the content of a constraint from its carrier of efficacy, StateGate achieves behavioral fidelity in long-chain tasks without losing the binding force of the prompt. Specifically, we first propose four independent compression strategies—deduplication, generalization, externalization, and text-to-gating—together with their corresponding decision chains, enabling principled prompt symbolization. Second, we design a three-component architecture consisting of a state file, a validator, and an orchestrator, which thoroughly separates judgment authority from execution authority. Third, we formally prove that the compliance of symbolic constraints is independent of context size. Finally, we validate the framework on a vulnerability-discovery task in the security domain; the results show that StateGate achieves 100\% execution closure and accurate vulnerability detection.
\end{abstract}


\section{Introduction}
\label{sec:intro}

With the rapid development of large language model (LLM) capabilities, LLM-centric agents have been widely adopted in cybersecurity, where long-horizon, multi-step security tasks such as autonomous vulnerability discovery, code auditing, and log analysis are typical application scenarios \cite{addis2026llm, ferrag2026prompt, he2026survey}. To make an agent execute a complex multi-step procedure completely, the prevailing practice is to encode the required domain knowledge, analysis workflow, and operational norms into a long prompt injected into the system context, in the hope that a complete methodology will guarantee proper execution. However, a long prompt does not necessarily produce complete execution. This problem poses three thorny challenges. First, context inflation and attention dilution: key instructions in the middle and later parts of a long window are easily diluted, causing the agent to skip or omit steps \cite{yu2025mitigate, hsieh2024found}. Second, uncontrolled multi-step execution: constraints rely on the agent's voluntary compliance, and obligations are often weakened into optional actions \cite{li2025let}. Third, high inference cost: call overhead grows with prompt length, constraining large-scale deployment \cite{sah2026verifier}. Together, these challenges point to a fundamental issue: writing a prompt in full is not equivalent to an agent doing everything.

Existing research has alleviated these problems from the directions of prompt compression, workflow orchestration, and behavior distillation, but its optimization objectives focus on semantic fidelity, process organization, or behavior substitution, and have not yet addressed behavioral fidelity. Prompt-compression methods pursue invariant information content before and after compression, but cannot guarantee that the key steps required after compression are not missing. Workflow-orchestration frameworks manage state transitions and task organization, but do not verify whether steps have actually been completed. Behavior-distillation and compilation methods replace model inference with executable programs, rather than constraining the model to execute a prescribed procedure completely in a long context. Although security-engineering practice has externalized constraints to deterministic runtimes, its goal is to enhance execution reliability and permission safety, not to serve prompt optimization, and it lacks a formal model and systematic strategies.

To address these challenges, we propose StateGate. Drawing on the idea of semantic constraint symbolization and exploiting enforcement by a deterministic runtime, StateGate achieves behavioral fidelity. Its rationale is that the content of a constraint and its carrier of effect can be separated. The content remains intact, while the carrier changes from text that is easily diluted to evidence-dependent runtime predicates. By externalizing key decision points into a state file, a validator, and an orchestrator, the binding force removed from the prompt is reconstructed in an external physical defense line, thereby eliminating the interference of attention mechanisms with instruction following at the root.

Specifically, the main contributions of this paper are as follows.
\begin{enumerate}[leftmargin=1.5em, itemsep=2pt, topsep=3pt]
	\item We establish a causal chain from constraint externalization to prompt compression to behavioral fidelity. We propose four independently attributable compression strategies---deduplication, generalization, externalization, and text-to-gating---which transform natural language into a symbolic language. We formally prove that the compliance rate of symbolic constraints is identically one, whereas that of semantic constraints decays as context length increases.
	\item Based on the symbolic language, we further propose a three-component architecture consisting of a state file, a validator, and an orchestrator. It thoroughly separates judgment authority from execution authority and fills the verification gap of ``whether a step is actually executed'' in existing schemes.
	\item We present a reusable open-source StateGate framework. With a three-layer form of declarative specification, generator, and replaceable backend, it generalizes text-to-gating into a general agent-engineering principle. Real-world experiments show that symbolic gating significantly outperforms baselines in stage compliance, coverage, and recall.
\end{enumerate}

The remainder of this paper is organized as follows. \Cref{sec:related} reviews related work. \Cref{sec:design} details the design and formal model of StateGate. \Cref{sec:theory} presents the theoretical analysis. \Cref{sec:experiments} reports the comparative experiments and performance analysis in a real security-audit scenario. \Cref{sec:conclusion} concludes the paper and discusses future directions.

\section{Related Work}
\label{sec:related}

Existing work can be divided into four directions according to its optimization objective: prompt compression, workflow orchestration and guardrails, behavior distillation and compilation, and constraint externalization in security engineering. These directions have made progress in semantic fidelity, process organization, behavior substitution, and permission safety, respectively, but none treats ``whether key steps are executed'' as a direct optimization objective. \Cref{tab:positioning} summarizes the positioning of existing methods and StateGate.

\begin{table}[t]
	\caption{Positioning of existing methods and StateGate.}
	\label{tab:positioning}
	\centering
	\footnotesize
	\begin{tabular}{@{}>{\raggedright\arraybackslash}p{0.17\textwidth}>{\raggedright\arraybackslash}p{0.21\textwidth}>{\raggedright\arraybackslash}p{0.25\textwidth}>{\raggedright\arraybackslash}p{0.27\textwidth}@{}}
		\toprule
		Research line & Representative work & Optimization objective & Boundary with StateGate \\
		\midrule
		Prompt compression & LLMLingua, Selective Context, Gist tokens & Semantic fidelity: no information loss after compression & Does not change the constraint carrier; behavioral fidelity is outside the objective \\ \\
		Workflow orchestration and guardrails & CoAct, LangGraph, Guardrails AI & Process organization: legal state transitions and compliant outputs & Within-stage completion relies on agent self-report; no artifact grounding \\ \\ 
		Behavior distillation and compilation & Skill-DisCo, agi compiler & Behavior substitution: fix behavior flows and bypass LLM inference & Inductive rather than symbolic; replaces execution instead of constraining it \\ \\
		Security-engineering constraint externalization & DeepSeek Harness, prompt-language, LLM-CTF-Solver, SANS & Execution reliability and permission safety & Externalization for hardening, not prompt reduction; point solutions without a compression recipe \\
		StateGate (this paper) & Four strategies + three components & Joint behavioral fidelity and prompt compression & Closed causal chain: externalize $\rightarrow$ compress $\rightarrow$ preserve behavior \\
		\bottomrule
	\end{tabular}
\end{table}

\subsection{Prompt Compression}
\label{sec:rw-compression}

Prompt-compression methods aim to preserve semantic information while reducing context length by deleting redundant text. For example, LLMLingua uses a small language model's perplexity to estimate token importance and achieves roughly 20$\times$ compression on question-answering tasks \cite{jiang2023llmlingua}; Selective Context retains the most relevant context fragments based on self-information \cite{li2023selective}; and Gist tokens encode long prompts into a small number of learnable soft prompts \cite{mu2023gist}. However, these methods are insufficient: compression changes text length, not the way constraints are carried. Compressed constraints remain in the context as natural language and are still subject to attention dilution. The presence of information is not equivalent to the execution of steps; these methods do not address the root cause of ``constraints being diluted,'' nor do they directly optimize execution metrics such as the step-skipping rate and diagnostic-chain completion rate.

\subsection{Workflow Orchestration and Guardrails}
\label{sec:rw-orchestration}

Workflow-orchestration and guardrail methods aim at legal state transitions and compliant outputs by explicitly structuring agent workflows as graphs or state machines. For example, CoAct employs a hierarchical collaboration architecture with a global planner and local executors \cite{hou2024coact}; LangGraph uses nodes, edges, and conditional edges to determine state-transition paths and provides checkpoint-based pause and resume \cite{langgraph}; and Guardrails AI applies structured validation and correction to generated outputs \cite{guardrailsai}. However, these methods still have shortcomings: orchestration frameworks provide a skeleton for state transitions, but leave whether a stage is truly completed to the agent's self-report. A legal graph transition is not equivalent to substantive step completion. Guardrails validate the compliance of output formats rather than adherence of the execution process to the methodology, and they lack artifact-level grounding.

\subsection{Behavior Distillation and Compilation}
\label{sec:rw-distillation}

Behavior-distillation and compilation methods aim to fix behavior flows and bypass LLM inference by converting successful execution trajectories into reusable programs. For example, Skill-DisCo distills successful agent trajectories into procedural skills encoded as parameterized finite-state machines \cite{guo2026skill}; the agi compiler compiles agent behavior into verified WebAssembly modules, so that process execution no longer depends on repeated LLM inference \cite{jaber2026agi}. However, these methods have a fundamental limitation: they follow a bottom-up inductive path that learns skills from known trajectories, whereas real tasks require agents to analyze unknown targets. Fixing behavior into a fixed workflow conflicts with the premise of autonomous discovery.

\subsection{Constraint Externalization in Security Engineering}
\label{sec:rw-externalization}

Security-engineering practice aims at execution reliability and permission safety by moving constraints from prompts to external runtimes. For example, DeepSeek Harness points out that when an agent can modify the codebase and read credentials, natural language can hardly constitute a security boundary \cite{deepseekharness}; prompt-language lets real command checks, rather than agent self-reports, determine whether a stage is complete \cite{promptlanguage}; LLM-CTF-Solver uses six-dimensional deadlock detection to trigger strategy switching \cite{llmctfsolver}; and the SANS guide emphasizes that skipping steps and accepting conclusions will not withstand verification \cite{wright2026sans}. However, these methods still have deficiencies: their externalization serves as a safety fuse and does not address the cognitive degradation caused by long contexts. They usually require retaining or even lengthening the original prompt and therefore cannot mitigate attention dilution. Moreover, they are mostly point solutions and lack a systematic prompt-compression mechanism and artifact-level grounding, so they cannot ensure that a multi-step methodology is truly executed in full.

In summary, none of the four directions answers the same question: after a prompt is condensed, are all the key steps that the agent should perform still performed? Existing externalization can prevent privilege escalation but not laziness; prompt compression can shrink text but cannot preserve behavior. This is precisely the core problem that this paper attempts to solve. StateGate absorbs the idea of constraint externalization and transforms semantic constraints into external symbolic gates. It fills the gap for a quantitative compression recipe and, with behavioral fidelity as its objective, unifies prompt optimization and execution completeness.

\section{StateGate Design}
\label{sec:design}

\subsection{Overview}
\label{sec:overview}

\begin{figure}[h]
	\centering
	\includegraphics[width=\textwidth]{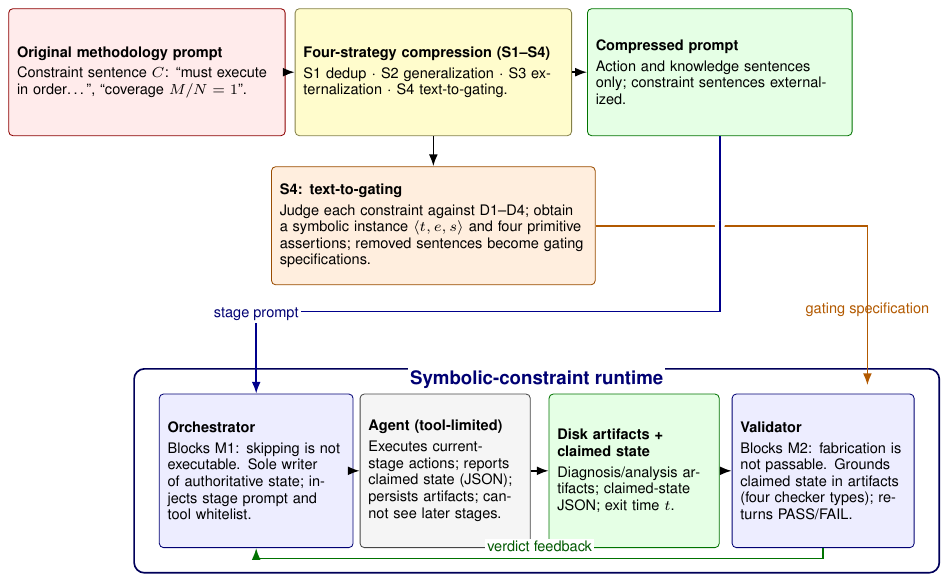}
	\caption{Overview of the StateGate methodology. The original methodology prompt is compressed by four strategies (S1--S4). S4 converts constraints into symbolic instances and gating specifications, which are enforced by the orchestrator--validator--state-file runtime.}
	\label{fig:overview}
\end{figure}

The output of StateGate consists of two parts: a symbolized methodology prompt and a symbolic-constraint runtime. Its core idea is simple: transform semantic constraints into symbolic constraints enforced by an external deterministic runtime. The prompt retains only ``what to do,'' while ``to what extent it must be done and what must not be skipped'' is physically intercepted by the runtime. \Cref{fig:overview} presents an overview of the method.

The rationality and feasibility of this design are not based on an empirical stacking of components but on rigorous theoretical derivation, detailed in \Cref{sec:theory}. Theoretical analysis shows that in a long context, the effective response probability of a semantic constraint decays exponentially with context size (Proposition~\ref{prop:semantic}); if judgment authority is transferred to an external deterministic runtime, the compliance rate is completely decoupled from context size (Proposition~\ref{prop:symbolic}). This conclusion provides solid support for the feasibility of StateGate: the only effective way to break instruction dilution caused by long contexts is to convert constraints from probabilistic responses inside the model into deterministic thresholds outside the model.

Accordingly, the scheme consists of two key components. Component~1, constraint symbolization, rewrites constraints in the prompt into runtime-decidable symbolic instances; it is carried by S4, with S1--S3 as pre-compression. Component~2, the gating loop, executes these symbolic instances during stage scheduling; it is carried jointly by the orchestrator, the validator, and the state file.

\subsection{Design Principles}
\label{sec:principles}

In a long context, execution degradation manifests in two forms: M1 (skipped invocation), where an action that should be executed is not executed, and M2 (fabricated result), where execution is claimed without supporting evidence. On this basis, we establish five design principles, summarized in \Cref{tab:principles}.

\begin{table}[h]
	\caption{StateGate design principles.}
	\label{tab:principles}
	\centering
	\footnotesize
	\begin{tabular}{@{}>{\raggedright\arraybackslash}p{0.20\textwidth}>{\raggedright\arraybackslash}p{0.44\textwidth}>{\raggedright\arraybackslash}p{0.28\textwidth}@{}}
		\toprule
		Principle & Meaning & Problem addressed \\
		\midrule
		P1 Constraint externalization & Key decision-point constraints move out of the prompt and are carried by a deterministic runtime & Makes compression possible \\ 
		P2 State separation & The authoritative state held by the orchestrator is strictly separated from the claimed state written by the agent & Prerequisite for anti-fabrication \\ 
		P3 Stage isolation & Each stage invokes an independent agent with only the current-stage prompt and the shared header & Prevents step skipping \\
		P4 Capability gating & A tool whitelist restricts the capabilities available at each stage & Prevents step skipping \\ 
		P5 Artifact grounding & The validator queries observable artifacts on disk instead of relying on agent self-reports & Prevents fabrication \\
		\bottomrule
	\end{tabular}
\end{table}

P1 is the guiding principle, and P2--P5 are its implementation conditions. Only when constraints are externalized and jointly carried by authoritative state, isolated execution, and grounded verification can the statement ``symbolic constraints cannot be bypassed'' hold at the mechanical level.

\subsection{Key Component 1: Constraint Symbolization}
\label{sec:symbolization}

As shown in \Cref{fig:overview}, S1--S3 are pre-compression strategies. S1 denotes deduplication, which removes repeated descriptions and examples; S2 denotes generalization, which abstracts device-specific descriptions into general statements; S3 denotes externalization, which moves declarative knowledge out of the prompt. These three strategies operate inside the text and change its length, but constraints still exist in natural-language form. S4 is fundamentally different: it changes the representational form of constraints, strips them from the prompt body, and rewrites them as symbolic instances outside the text.

The core motivation for this transformation is that the two carriers are not mechanically equivalent. If they were equally effective, moving a constraint out would amount to canceling it. A constraint carrier is characterized by its judgment subject and evidence source. A constraint expressed in natural language whose validity can only be reflected by the agent's response behavior is called a \emph{semantic form}; a constraint expressed as a predicate whose validity is judged by a deterministic runtime based on observable evidence is called a \emph{symbolic form}. The essential difference lies in the implementation mechanism: the semantic form fulfills a probabilistic obligation and, in scenarios with many constraints and a long context, is highly susceptible to failure because attention allocation is diluted; the symbolic form provides a deterministic threshold whose effectiveness is completely independent of context size. Therefore, improving constraint effectiveness cannot rely merely on ``writing it more prominently.'' Instead, a constraint must be converted from ``text that the model must remember'' into ``a fact that the runtime must satisfy.'' This is the fundamental starting point of S4.

However, not every natural-language constraint is eligible for conversion. Forced conversion would silently discard undecidable semantics. To this end, S4 defines a four-question decision chain that delimits the conversion boundary, as shown in \Cref{tab:decision}.

\begin{table}[h]
	\caption{The four decision questions for constraint symbolization.}
	\label{tab:decision}
	\centering
	\footnotesize
	\begin{tabular}{@{}>{\raggedright\arraybackslash}p{0.21\textwidth}>{\raggedright\arraybackslash}p{0.40\textwidth}>{\raggedright\arraybackslash}p{0.31\textwidth}@{}}
		\toprule
		Decision question & Decision rule & Counterexample \\
		\midrule
		D1 Decidability & Mechanically decidable truth semantics exist & Aesthetic or trade-off soft constraints \\ 
		D2 Evidence availability & Satisfaction is provable by persistent artifacts & Purely internal reasoning without persistent evidence \\ 
		D3 Timing of validity & The constraint crosses a stage boundary & A one-off requirement within a stage \\ 
		D4 Deletion test & The normative semantics are fully carried by the symbolic instance & Statements that also serve as knowledge background \\
		\bottomrule
	\end{tabular}
\end{table}

These four questions are not a random list. They characterize the four thresholds that a constraint must cross before it can ``leave the LLM and be executed by code.'' First, can the LLM identify an objective criterion (D1: set the standard)? Second, can the program find evidence on disk to verify it (D2: find the evidence)? Third, at which checkpoint in the process should this verification action be inserted (D3: set the position)? Fourth, after deleting the original sentence, will the system lose the original semantic constraint (D4: preserve the bottom line)? Only when all four steps pass has the constraint truly completed a safe externalization. If any question fails, the constraint remains in the prompt body, thereby ensuring behavioral fidelity of the overall compression.

A constraint that passes the decision chain is abstracted into a formal triple $\langle t,e,s\rangle$, where $t$ is the check time determined by D3, $e$ is the grounded evidence determined by D2, and $s$ is the symbolic form.

\begin{figure}[H]
	\centering
	\includegraphics[width=\textwidth]{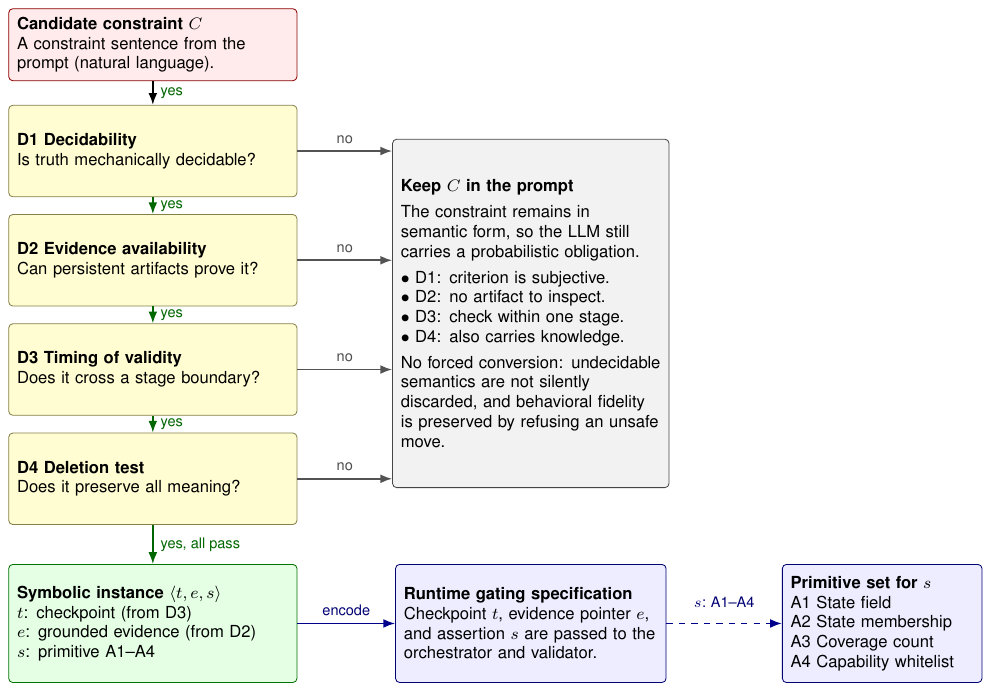}
	\caption{Constraint symbolization architecture. Each candidate constraint passes the D1--D4 decision chain and, if eligible, is converted into a symbolic instance $\langle t,e,s\rangle$, where $s$ selects one of the four primitives A1--A4.}
	\label{fig:symbolization}
\end{figure}

Regarding the symbolic form $s$, the system must guarantee completeness so as to cover all decidable constraint scenarios. In formal verification, an objective assertion about system state can usually be reduced to four basic dimensions: single-point fact, set membership, cardinality, and capability permission. Based on these four dimensions, we map $s$ to four mutually exclusive and complete basic primitives; their definitions and mappings are given in \Cref{tab:primitives}.

\begin{table}[h]
	\caption{The four basic primitives of symbolic form.}
	\label{tab:primitives}
	\centering
	\footnotesize
	\begin{tabular}{@{}>{\raggedright\arraybackslash}p{0.23\textwidth}>{\raggedright\arraybackslash}p{0.17\textwidth}>{\raggedright\arraybackslash}p{0.52\textwidth}@{}}
		\toprule
		Primitive type & Verification dimension & Meaning and decision logic \\
		\midrule
		A1 State-field assertion & Single-point fact & Read the field values of a specific artifact file and decide whether preset conditions are satisfied \\ 
		A2 State-machine membership assertion & Set membership & Obtain the current runtime state and decide whether it belongs to a preset legal state set \\ 
		A3 Coverage-count assertion & Cardinality & Count the artifacts in a specified directory or array and decide whether a threshold is reached \\ 
		A4 Capability-whitelist assertion & Capability permission & Inspect the tool-call log or orchestrator authorization records and decide whether a privilege violation occurred \\
		\bottomrule
	\end{tabular}
\end{table}

At the underlying data-structure level, these four primitives precisely correspond to the core requirements of flow control, progress verification, and permission management. Any hard constraint that passes D1 can be losslessly reduced to one of these four primitives, thereby ensuring expressiveness while avoiding unnecessary primitive redundancy. The complete symbolization architecture is shown in \Cref{fig:symbolization}.

At this point, the problem of how to transform semantic constraints into the representational form of symbolic constraints has been solved. It should be emphasized that the overall compression process formed by S1--S4 is not arbitrary text deletion. To prove its rigor, we establish three core criteria for compression attributability: orthogonality of provenance, independence of measurement, and reproducibility of process. These criteria not only ensure that compression results are interpretable and reviewable, but also lay the theoretical foundation for \Cref{sec:theory}, where we prove that ``compression does not harm fidelity.'' How to ensure that the generated symbolic assertions are strictly enforced at runtime is developed in the next key component, the gating loop.

\subsection{Key Component 2: The Gating Loop}
\label{sec:gating}

Generating symbolic assertions is not equivalent to realizing constraint effectiveness. If an LLM agent is allowed to check its own compliance, complex tasks are highly prone to fabricated results (M2) and skipped steps (M1). To this end, we design a three-component architecture consisting of a state file, a validator, and an orchestrator. Its core logic is to thoroughly separate ``judgment authority'' from ``execution authority.'' The responsibility boundaries and defense mechanisms of the three components are summarized in \Cref{tab:components}.

\begin{table}[h]
	\caption{The three-component model.}
	\label{tab:components}
	\centering
	\footnotesize
	\begin{tabular}{@{}>{\raggedright\arraybackslash}p{0.17\textwidth}>{\raggedright\arraybackslash}p{0.42\textwidth}>{\raggedright\arraybackslash}p{0.31\textwidth}@{}}
		\toprule
		Component & Core permission restricted & Defense objective \\
		\midrule
		State file & Blocks the agent's self-certification authority & Provides the data object for checking M2 \\
		Validator & Blocks the agent's judging authority & Intercepts fabricated results (M2) \\
		Orchestrator & Blocks the agent's authority to escalate privileges and skip steps & Intercepts skipped invocations (M1) \\
		\bottomrule
	\end{tabular}
\end{table}

These three components are not a simple functional aggregation. Rather, they build complementary physical defense lines against the two failure modes, as described below.

\paragraph{First defense line: ``claim--verification'' separation against M2 (fabricated results).}
To cut off the possibility that the agent acts as ``both athlete and referee,'' the scheme splits state into two isolated subjects. The first is the claimed state $s_{\mathrm{claim}}$, written by the agent and representing its subjective progress. The second is the authoritative state $s_{\mathrm{auth}}$, maintained exclusively by the orchestrator, to which the agent has no write permission. The dedicated validator recognizes only real artifacts on disk, not the agent's self-check report. By greatly raising the time cost required to fabricate artifacts, this mechanism makes ``faking'' unprofitable.

\paragraph{Second defense line: ``physical-level isolation'' against M1 (skipped invocations).}
Artifact verification alone cannot prevent the skipping behavior of ``submitting without doing the work.'' Therefore, during stage scheduling, the orchestrator injects only the current-stage prompt and issues only the current-stage whitelisted tools. The agent cannot see instructions for later stages and cannot invoke tools for later stages, thereby physically cutting off the possibility of skipping. At the code-execution layer, these three defense lines operate together through the orchestrator's scheduling main loop, shown in \Cref{alg:orchestrator}.

\begin{figure}[h]
	\centering
	\includegraphics[width=\textwidth]{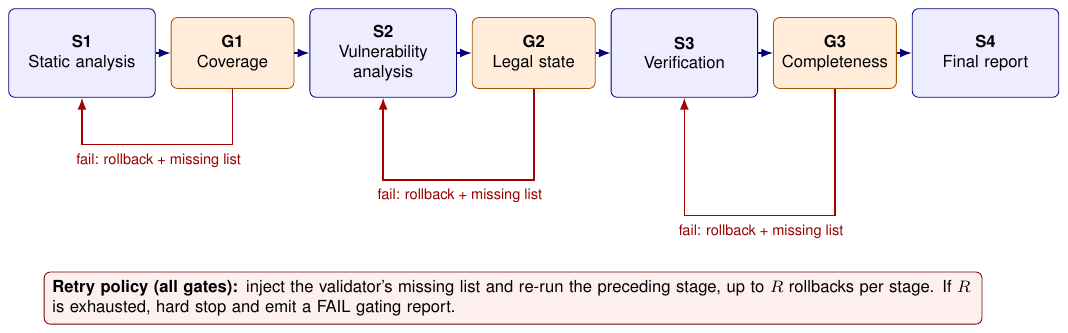}
	\caption{Gated state-machine loop. G1--G3 are boundary checkpoints enforcing coverage (P3), legal state membership (P2), and diagnostic-chain completeness (P1), respectively; a failed gate triggers rollback with the missing list or, after $R$ retries, a hard stop.}
	\label{fig:gating}
\end{figure}

In terms of execution timing, the two defense lines are connected into a rigorous control flow by the orchestrator main loop. As shown in \Cref{fig:gating}, the system places three typical gating checkpoints at critical points of the process. G1, the coverage gate, guards the exit of the static-analysis stage and checks the number of files persisted to disk to prevent the agent from finishing early before the target has been fully examined. G2, the state-machine gate, guards the stage entrance and checks the authoritative state snapshot to prevent the agent from illegally advancing the process. G3, the completeness gate, guards the entrance of final verification and checks the evidence files for each step of the diagnostic chain to prevent the agent from omitting key steps and outputting a conclusion directly. These three checkpoints correspond to the three core primitives of coverage counting, state-machine membership, and state field, respectively.

\begin{algorithm}[h]
	\caption{Orchestrator main loop: stage scheduling and gated transitions.}
	\label{alg:orchestrator}
	\begin{algorithmic}[1]
		\Require Stage sequence $S_1,\ldots,S_n$ (with section anchors and tool whitelists); attachment table of boundary checkpoints $G_j$; maximum retry count $R$.
		\Ensure Gating report.
		\State $\mathit{state} \gets \textsc{InitAuthState}(\mathit{scan\_id}, \mathit{current}=S_1, \mathit{gates}=\emptyset, \mathit{rollback}=0)$
		\For{$i \gets 1$ \textbf{to} $n$}
			\If{$\mathit{state.current}$ has passed $S_i$}
				\State \textbf{continue} \Comment{resume scan; do not re-verify passed gates}
			\EndIf
			\State $\mathit{prompt} \gets \textsc{SharedHeader} \oplus S_i.\mathit{section}$ \Comment{stage isolation: no later-stage content}
			\State $\mathit{tools} \gets \textsc{Whitelist}(S_i)$ \Comment{capability gating}
			\Repeat
				\State $\textsc{RunAgent}(\mathit{prompt}, \mathit{tools})$ \Comment{one independent execution}
				\State $\mathit{verdict} \gets \textsc{Validate}(\textsc{ExitCheckpoint}(S_i), \mathit{claimed\_state}, \mathit{stage\_artifact\_dir})$
				\If{$\mathit{verdict} = \textsc{PASS}$}
					\State \textbf{break}
				\EndIf
				\State $\mathit{prompt} \gets \mathit{prompt} \oplus \textsc{MissingList}(\mathit{verdict.missing})$ \Comment{rollback: inject missing items}
				\State $\mathit{rollback}[i] \gets \mathit{rollback}[i] + 1$
			\Until{$\mathit{rollback}[i] \geq R$}
			\If{$\mathit{verdict} \neq \textsc{PASS}$}
				\State \textbf{hard stop}; output gating report; \Return \textsc{FAIL}
			\EndIf
			\State $\mathit{state} \gets \textsc{Advance}(S_{i+1})$ and immediately persist \Comment{anchor for recovery and audit}
		\EndFor
	\end{algorithmic}
\end{algorithm}

\subsection{Framework Implementation}
\label{sec:framework}

If the mechanisms described in the previous subsections were transferred to new tasks through manual modification, the process would be time-consuming and the results irreproducible. We therefore solidify them into an open-source framework, StateGate. Its core design principle is ``single source of truth'': all task knowledge converges into one declarative specification, and both prompts and validators are derived artifacts. The framework has three layers.

\paragraph{Specification layer (the only input).}
The specification file stores no prompt text. It declares only six elements, including stage partitioning, gate attachment, and tool issuance. Only the ``gating rules'' depend on LLM judgment; the remaining elements are either provided by the author or mechanically projected by rules, which minimizes the scope of subjective intervention.

\paragraph{Generation layer (deterministic compilation).}
To ensure artifact consistency, the generation layer reads only the specification and not the source files. Stage prompts and validators are generated in a rule-driven manner, ensuring that repeated generation from the same specification is byte-identical. In addition, the generation layer includes a ``targeted sabotage self-check'': by actively tampering with simulated artifacts, it forces the corresponding gate to report an error, thereby effectively identifying and eliminating ``always-true gates,'' i.e., invalid gates that never report errors.

\paragraph{Runtime layer (replaceable engine).}
The state-file and validator logic are fixed, while the underlying execution backend is replaceable. However, there is one non-negotiable hard constraint: the backend must support tool-level capability restriction. The physical basis for intercepting M1 is the ``physical absence of tool invocation.'' If the backend can only impose soft, prompt-level restrictions, capability gating degenerates into an ineffective natural-language sentence, and the defense line collapses.

Through this three-layer decoupling, StateGate turns the ``text-to-gating'' methodology into a reusable and auditable engineering pipeline.

\section{Theoretical Analysis}
\label{sec:theory}

This section provides formal theoretical support for StateGate. We first model the attributability of the compression strategies described above. We then construct a system model of multi-step workflows. Finally, we derive propositions on the evolution of compliance under different constraint forms.

\subsection{Attributability of Compression}
\label{sec:attributability}

Any method that shortens long prompts faces the same question: does the reduction come from principled operations or from a one-off trade-off? The problem with the latter is not its magnitude but its irreproducibility. This subsection provides a set of criteria independent of specific text to determine whether a compression strategy is attributable. We regard a prompt as a sequence of knowledge units $T=\langle u_1,u_2,\ldots,u_n\rangle$, where each unit is a statement. The four compression strategies correspond to processing functions $S_1$ through $S_4$, and the overall compression effect is their sequential composition:
\begin{equation}
	S = S_4 \circ S_3 \circ S_2 \circ S_1,
	\label{eq:composition}
\end{equation}
where $S_1$ is applied first and $S_4$ last.

To prove that the process is attributable, three criteria must be satisfied.

\paragraph{Orthogonality of provenance.}
Each knowledge unit may be processed for at most one reason. Let the strategy set be $U=\{0,1,2,3,4\}$, and define the attribution map $\alpha: T \rightarrow U$. For any unit $u$ in the sequence, $\alpha(u)=i$ means that $u$ is attributed to strategy $i$, and $\alpha(u)=0$ means that $u$ is retained. Let $A_i=\{u\in T:\alpha(u)=i\}$. Orthogonality is then expressed as
\begin{equation}
	A_i \cap A_j = \emptyset \quad (i \neq j), \qquad \bigcup_{i=1}^{4} A_i \subseteq T.
	\label{eq:orthogonality}
\end{equation}
\Cref{eq:orthogonality} establishes formal non-overlap. Its substance lies in the objectivity of the attribution rules. The map $\alpha$ must be determined solely by properties of the text itself: duplicate detection for S1, generalized expression for S2, declarative knowledge for S3, and externalizable constraints for S4. The four feature classes are mutually exclusive, so the attribution is not subject to subjective preference. Without objective rules, a third party cannot reproduce the same mapping.

\paragraph{Independence of measurement.}
The contribution of each compression step S1--S4 must be measurable separately. Let the initial text before compression be $T^{(0)}=T$, and let the text after the first $i$ steps be $T^{(i)}=S_i(T^{(i-1)})$. For any unit, let $\delta_i(u)$ be the change in the number of lines caused by the $i$-th reason. Since the scopes of the strategies do not overlap, the reduction of each step can be computed directly as $\Delta_i=\sum_{u\in T}\delta_i(u)$. The step reduction rate $\rho_i$ and the total compression rate $\rho$ are then
\begin{equation}
	\Delta_i = \sum_{u\in T}\delta_i(u), \qquad
	\rho_i = \frac{\Delta_i}{|T^{(0)}|}, \qquad
	\rho = \sum_{i=1}^{4}\rho_i = 1 - \frac{|T^{(4)}|}{|T^{(0)}|}.
	\label{eq:measurement}
\end{equation}
The physical meaning of \Cref{eq:measurement} is clear: the total compression rate equals the sum of the reduction rates of the four independent strategies and exactly equals the size difference between the initial and final texts.

\paragraph{Reproducibility of process.}
Each operation must be a deterministic mapping without random rewriting. All four decision criteria can be fully characterized by objective rules. When the rules and parameters are public, any third party can recompute exactly the same reduction rate from the same rules.

We have thus shown that the reduction of prompt length is absolutely safe and traceable. However, reducing text is only the first step. The removed constraints must continue to take effect in some form within the system. This requires shifting our perspective from the text level to the execution level and systematically modeling multi-step workflows.

\subsection{Formal System Model}
\label{sec:system-model}

To achieve quantitative analysis from ``text operations'' to ``system operation,'' we first rigorously define multi-step tasks and execution metrics.

\paragraph{Notation.}
Let $T$ be the prompt text sequence and $C$ the context size. Let $R$ be the set of required transitions in the workflow, and let $m=|R|$ be the number of constraints. Let $\varphi(C)$ be the compliance function, $q(\ell,C)$ the probability that a single semantic constraint statement $\ell$ is effectively responded to, and $\delta$ the failure probability of a single judgment step.

\begin{definition}[Workflow model]
\label{def:workflow}
A multi-step agent workflow is denoted by $W=(S,\rightarrow,P)$, where $S$ is the set of stage states, $\rightarrow \subseteq S\times S$ is the allowed transition relation, and $P=\{p_s\}_{s\in S}$ is a family of completion predicates, with $p_s:E\rightarrow\{0,1\}$ deciding whether stage $s$ reaches its completion condition based on evidence $e$ in the observable evidence space $E$.
\end{definition}

\begin{definition}[Compliance]
\label{def:compliance}
Let $R_{\mathrm{req}}$ be the set of transitions required by the task specification, and let $R\subseteq \rightarrow$ (called the required transitions). The compliance rate is defined as the probability that all these required transitions are successfully completed in a single run:
\begin{equation}
	\varphi(C) = \Pr\left[R^{*}\right],
	\label{eq:compliance}
\end{equation}
where $R^{*}$ denotes the event that ``the completion predicates of all transitions in $R$ are judged satisfied at the corresponding stage exits.'' This probability can be estimated directly from the observed completion frequency in actual runs.
\end{definition}

The compliance rate measures execution completeness at the transition level. With this metric in hand, we next introduce the objective physical limit of long contexts and derive the consequences of this limit under different constraint carriers.

\subsection{Theoretical Assumptions and Derivations}
\label{sec:derivations}

To analyze the change in $\varphi$, we must introduce a premise of attention decay in long-context environments. We make the phenomenon of attention decay explicit as Assumption~\ref{ass:dilution} (attention dilution), which has the following properties.

\begin{assumption}[Attention dilution]
\label{ass:dilution}
For a given statement $\ell$ and a context of size $C$, let $q(\ell,C)$ be the probability that the statement is effectively responded to and adopted by the agent. Then:
\begin{enumerate}[leftmargin=1.5em, itemsep=1pt, topsep=2pt]
	\item \emph{Monotonicity:} $q(\ell,C)$ is monotonically non-increasing in $C$.
	\item \emph{Negligibility:} under a fixed attention budget and without positional bias, $\lim_{C\rightarrow\infty} q(\ell,C)=0$. That is, as the context size tends to infinity, the probability of effective response tends to zero.
\end{enumerate}
\end{assumption}

Based on the above definitions and assumption, when multiple constraints act together, the overall compliance is a composition of their individual response outcomes. Under the semantic and symbolic forms, this composition exhibits fundamentally different mathematical properties.

\begin{proposition}[Compliance decay under the semantic form]
\label{prop:semantic}
If the release of a transition is determined by the semantic form of its constraint, then under the independence assumption, the semantic compliance rate is
\begin{equation}
	\varphi_{\mathrm{sem}}(C) = \prod_{r\in R} q_r(C) \;\leq\; \bar{q}(C)^{m},
	\qquad
	\bar{q}(C) = \max_{r\in R} q_r(C).
	\label{eq:semantic}
\end{equation}
\end{proposition}

In \Cref{eq:semantic}, $\prod_{r\in R}q_r(C)$ is the product of the effective response probabilities of the individual constraints, $\bar{q}(C)$ is the maximum probability, and $m=|R|$ is the number of required transitions. The equation shows that, under the independence assumption, the compliance rate decays exponentially with the number of constraints $m$. Moreover, as $C\rightarrow\infty$, the maximum response probability $\bar{q}(C)\rightarrow 0$, so that $\varphi_{\mathrm{sem}}(C)\rightarrow 0$. This reveals the essential fragility of natural-language constraints in long contexts.

\begin{proposition}[Compliance invariance under the symbolic form]
\label{prop:symbolic}
If judgment authority is externalized and state advancement is restricted, the symbolic compliance rate is
\begin{equation}
	\varphi_{\mathrm{sym}}(C) = (1-\delta)^{m}.
	\label{eq:symbolic}
\end{equation}
In the idealized case where $\delta=0$, $\varphi_{\mathrm{sym}}\equiv 1$, independent of both $C$ and $m$. This proves that symbolic constraints cut off the negative effect of context length on compliance.
\end{proposition}

Proposition~\ref{prop:symbolic} reveals the ideal effectiveness of symbolic constraints, but its validity depends heavily on the premise that ``judgment authority belongs entirely to the runtime.'' In the real physical world, not all constraints can perfectly satisfy this premise.

\subsection{Boundary Analysis and Corollary}
\label{sec:boundary}

If judgment authority is returned to the model, the symbolic form degrades to the semantic form in terms of actual effectiveness. More importantly, in real tasks, not all constraints are groundable. For constraints that cannot be proven, judgment authority can only remain on the model side. Therefore, the required transition set $R$ is partitioned as a whole into two independent subsets: provable transitions $R_{\mathrm{sym}}$ and unprovable transitions $R_{\mathrm{sem}}$.

\begin{corollary}[Graceful degradation]
\label{cor:graceful}
Combining the two types of constraints, the overall compliance rate of the system is
\begin{equation}
	\varphi(C) = (1-\delta)^{|R_{\mathrm{sym}}|}\cdot\varphi_{\mathrm{sem}}(R_{\mathrm{sem}},C),\varphi_{\mathrm{sem}}(R_{\mathrm{sem}},C)=\prod_{r\in R_{\mathrm{sem}}}q_r(C)
	\label{eq:graceful}
\end{equation}
\end{corollary}

In the ideal state, the overall compliance of the system is entirely determined by the constraints that cannot be grounded. This means that the essence of ``graceful degradation'' is as follows: when some constraints cannot be symbolized, the cost of maintaining system fidelity is no longer reflected in the loss of symbolic compliance, but is shifted to a decrease in the compression rate. The more constraints remain in the prompt body, the weaker the compression effect, but the underlying fidelity capability of the system remains solid. Thus, the engineering intuition that ``the compression rate drops, but fidelity is not compromised'' receives rigorous formal support.

\section{Performance Analysis and Simulation Experiments}
\label{sec:experiments}

This section evaluates, through controlled experiments in a real environment, the ability of the StateGate framework to guarantee the completeness of LLM-agent execution behavior. The core goal of StateGate is not to replace the agent's own reasoning ability, but to ensure that a long and complex task methodology is followed stage by stage without omission.

To verify this goal, we select a long-chain, high-complexity security-audit task. In such tasks, the agent is highly prone to attention drift in a long context, leading to behavioral degradation such as skipping stages or fabricating conclusions. We use ``whether the deep vulnerability can be accurately discovered in the end'' as the core validation metric for ``whether the execution behavior is faithful.'' The trigger path of the target vulnerability is highly concealed. The agent must first execute all required audit stages in full---target extraction, decompilation, full triage, and logic closure---without omitting any link; only then can the vulnerability be successfully detected. Accordingly, we set up four comparison schemes and conduct a quantitative analysis from three dimensions: coverage, execution process, and final result. (Note: an ablation study of the compression strategies will be added in a follow-up version.)

\subsection{Experimental Setup}
\label{sec:setup}

We conduct controlled experiments in a real environment to evaluate the ability of StateGate to guarantee LLM-agent execution completeness. The core objective of StateGate is to ensure that a long methodology is executed stage by stage without omission, rather than to replace the model's own reasoning ability.

We select a long-chain, high-complexity security-audit task. The task hides a deep authentication-bypass vulnerability, and its detection conditions are extremely stringent: the agent must fully execute all stages, including target extraction, decompilation, full triage, and logic closure. Skipping or omitting any link will cause a missed detection. Therefore, ``whether the vulnerability can be precisely detected'' becomes the strictest touchstone for ``whether the execution behavior is faithful.''

The four schemes are: (i) vanilla scan, which uses a single long prompt without external orchestration or gating; (ii) LangGraph-plain, which organizes the workflow as a graph but does not enforce artifact-level gating; (iii) LangGraph-gated, which adds stage-level gating to the graph but still relies on the agent's self-report for within-stage completion; and (iv) StateGate, the full framework proposed in this paper.

\subsection{Overall Experimental Results and Analysis}
\label{sec:overall-results}

\Cref{tab:performance} summarizes the core performance metrics of the four schemes. The data reveal a counterintuitive phenomenon: the scheme that generated the largest number of artifacts did not achieve the highest execution completeness.

\begin{table}[h]
	\caption{Core performance comparison of the four schemes.}
	\label{tab:performance}
	\centering
	\small
	\begin{tabular}{@{}lcccc@{}}
		\toprule
		Evaluation dimension & Vanilla scan & LangGraph-plain & LangGraph-gated & StateGate \\
		\midrule
		Total time & 34.7 min & 55.1 min & 79.8 min & 66.5 min \\
		Decompiled Java files & 6 & 0 & 25,232 & 178 \\
		Java files entering analysis & 6 & 0 & 19 & 178 \\
		Java triage rate & 100\% (tiny denominator) & n/a & 0.075\% & 100\% \\
		Target-class reach stage & 2 / 5 & 1 / 5 & 3 / 5 & 5 / 5 \\
		Target vulnerability confirmed & Missed & Missed & Missed & Confirmed \\
		\bottomrule
	\end{tabular}
\end{table}

The macroscopic data reveal two typical failure modes of existing schemes in terms of execution completeness.

The first is ``material-processing deficit.'' The vanilla scan takes the shortest time, but without mandatory guidance, the agent decompiles only a very small number of files after extracting the target class files, missing the core target class.

The second is ``insufficient cognitive investment.'' Although LangGraph-gated uses graph orchestration to force the generation of 25,232 decompiled files, only 19 of them actually enter the analysis scope, yielding a triage rate of only 0.075\%. Consequently, it fails to hit the vulnerability.

In contrast, StateGate generates 178 files and achieves a 100\% analysis triage rate. It not only completes the five-level reachability chain in full but also successfully confirms the target vulnerability, achieving behavioral fidelity while maintaining execution efficiency.

\subsection{Coverage Analysis}
\label{sec:coverage}

To characterize the above failure mechanisms, we introduce a ``five-level target-class reachability funnel,'' shown in \Cref{tab:funnel}. The funnel divides vulnerability discovery into five progressive stages, from ``file existence'' to ``logic closure.''

\begin{table}[h]
	\caption{Five-level target-class reachability funnel.}
	\label{tab:funnel}
	\centering
	\small
	\begin{tabular}{@{}lcccc@{}}
		\toprule
		Evaluation dimension & Vanilla scan & LangGraph-plain & LangGraph-gated & StateGate \\
		\midrule
		Target class exists in dependency package & $\checkmark$ & $\checkmark$ & $\checkmark$ & $\checkmark$ \\
		Corresponding class file extracted & $\checkmark$ & $\times$ & $\checkmark$ & $\checkmark$ \\
		Decompiled into readable source artifact & 7\% & 0\% & 100\% & 100\% \\
		Report opens/references target class & 0\% & 0\% & 0\% & 100\% \\
		Authentication-chain logic closure & No & No & No & Yes \\
		\bottomrule
	\end{tabular}
\end{table}

The data delineate the execution blind spots of each scheme. LangGraph-plain stops at stage two and never enters the decompilation surface. The vanilla scan stops at stage three. Although LangGraph-gated persists source code to stage three, its report citation rate is 0\%. Only StateGate crosses stages four and five.

This directly confirms the core condition of behavioral fidelity: ``decompilation persisted to disk'' is only a necessary, not a sufficient, condition for vulnerability discovery; ``the report is opened and traced to the decision point'' is the sufficient condition for the vulnerability to be successfully detected. StateGate opens the last link from physical files to cognitive closure precisely through external mandatory gating.

\subsection{Experimental Summary}
\label{sec:exp-summary}

Combining the four-scheme comparison and the five-level funnel data, this experiment reveals the mechanism of long-chain agent execution failure and the core value of StateGate. Pure workflow orchestration cannot guarantee behavioral fidelity. The self-referential validation it relies on has a structural blind spot: the system can only verify ``what the agent claims to have done,'' but cannot discover ``what the agent never listed.'' In contrast, the key to StateGate's fidelity is the ``thorough externalization of judgment authority.'' By transforming the semantic form into the symbolic form and deriving the three components of state file, validator, and orchestrator, StateGate moves the judgment standard from the agent to an external deterministic runtime. Experiments show that only by forcing the execution boundary through an external physical defense line and compelling the agent to complete evidence collection in response to gate errors can cognitive laziness and material omission be truly eliminated, achieving 100\% deep closure and precise vulnerability detection.

\section{Conclusion and Discussion}
\label{sec:conclusion}

This paper addresses the behavioral degradation of long-chain LLM agents in complex tasks, where ``prompts are written in full but steps are not executed in full,'' and proposes StateGate, a behavioral-fidelity prompt-optimization method. The core contribution is to break through the limitation of traditional methods that perform text compression only at the level of ``semantic fidelity,'' advancing the optimization objective to ``behavioral fidelity.'' Specifically, by constructing a causal chain of ``constraint externalization $\rightarrow$ prompt compression $\rightarrow$ behavioral fidelity,'' StateGate transforms semantic constraints into symbolic constraints decidable by a deterministic runtime. Combined with the four strategies of deduplication, generalization, externalization, and text-to-gating, it achieves a win--win outcome of prompt reduction and execution completeness in real long-chain tasks such as security auditing.

Experimental results show that the methodological characteristic of StateGate is that it does not attempt to persuade the model to work harder through ``better prompts.'' Instead, through the physical isolation and grounded verification of an ``external contract,'' it forcibly constrains the execution boundary of the agent. The data from the four controlled experiments demonstrate that only by transforming constraints in the methodology into runtime-enforced judgments can step skipping and omission in long tasks be thoroughly eliminated, truly achieving behavioral fidelity for long-chain agents.

Although StateGate has achieved significant results in guaranteeing execution completeness, one boundary remains to be broken in practical application and theoretical derivation: the framework currently mainly addresses the fidelity of ``execution behavior'' and cannot directly intervene in or improve the ``substantive reasoning quality'' of the model's outputs. In other words, an external contract can force the agent to consult the target source code and generate an analysis report, but if the model itself has weak code-understanding ability, it may still reach an incorrect conclusion. Although this is essentially a capability bottleneck of the underlying LLM, engineering and methodological means still leave room to mitigate this limitation. In future research and framework iteration, two directions can be explored.

\begin{enumerate}[leftmargin=1.5em, itemsep=2pt, topsep=3pt]
	\item \emph{Building a cross-validation mechanism.} Introduce multi-agent collaboration or a ``red-team/blue-team'' mode at key judgment nodes. For example, one model generates vulnerability hypotheses while another independent model plays the role of reviewer, using external tools to logically collide the artifacts, thereby hedging the cognitive bias of a single model.
	\item \emph{Knowledge augmentation and tool extension.} Convert domain-specific expert experience into ``deterministic verification predicates'' similar to those in StateGate, so that when facing complex code beyond its training distribution, the model can rely on external toolchains and knowledge graphs to perform substantive causal reasoning rather than mere surface text matching.
\end{enumerate}

In summary, StateGate provides a solid engineering foundation for the reliable execution of long-chain agents. How to further integrate ``constraint of behavior'' with ``augmentation of cognition'' will be key to realizing fully autonomous, highly trustworthy agents.

 ---------------------------------------------------------------------------

\end{document}